# New Heuristic Rounding Approaches to the Quadratic Assignment Problem

Wajeb Gharibi[1] and Yong Xia[2]

1. Computer Science & Information Systems College, Jazan University, Jazan P.O.Box 2096, KSA
2. LMIB of the Ministry of Education, School of Mathematics and System Sciences, Beihang University, Beijing 100191, China

**Abstract:** Quadratic assignment problem is one of the great challenges in combinatorial optimization. It has many applications in Operations research and Computer Science. In this paper, the author extends the most-used rounding approach to a one-parametric optimization model for the quadratic assignment problems. A near-optimum parameter is also predestinated. The numerical experiments confirm the efficiency.

## 1. Introduction

The quadratic assignment problem (QAP) is one of the great challenges in combinatorial optimization. It is known to be NP-hard. An $\varepsilon$-solution is NP-hard too. Formally the QAP can be written as

$$QAP : \min f(x) = tr(AXB + C)X^T \quad (1)$$
$$s.t. \quad X \in \Pi \quad (2)$$

where, A, B and C are $n \times n$ matrices, $tr$ denotes the trace of a matrix, and $\Pi$ is the set of $n \times n$ permutation matrices. Throughout we assume A and B are symmetric. For comprehensive survey of QAPs, please refer to Refs. [1-5].

In this article we focus on discussing rounding approach. It is most used in many heuristic procedures such as the scatter search [6], the continuation method [7] and the approach based on solving relaxed quadratic programming [8]. In many cases an infeasible solution $X_C$ is obtained from the combination of some feasible points or a / an (local) optimal solution of the relaxed problem. We have to generate a high-quality feasible point $X \in \Pi$ using $X_C$.

The stand way to reach the goal is to find out $X$ which is the closest to $X_C$, i.e., we have to solve

$$\min tr(X - X_C)(X - X_C)^T \quad (3)$$
$$s.t. \quad X \in \Pi \quad (4)$$

Since $XX^T = I$, it reduces to the following linear assignment problem (LAP):

$$\max tr X_C X^T \quad (5)$$
$$s.t. \quad X \in \Pi \quad (6)$$

We notice that the linear assignment problem (5) – (6) can be exactly solved in $O(n^3)$. This optimal solution is denoted by $X_0$. Other speed-up approaches [6, 8] are based on inexactly solving (5) – (6). They are also applicable in our cases. Without lose of generality, in this article we only concern the exact method solving LAP.

Our paper is organized as follows: In section 2 we present the one-parametric optimization model. In section 3 we deduce a high-quality parameter. Numerical results are presented in section 4. Concluding remarks are made in section 5.

**Notation.** The Kronecker product of matrix A and B

---

**Corresponding author:** Wajeb Gharibi, associate professor, Ph.D., research fields: computer science and operations research. E-mail: gharibiw2002@yahoo.com.

Yong Xia, assistant professor, Ph.D., research fields: optimization.

is denoted $A \otimes B$. $Diag(a)$ is a diagonal matrix with diagonal components $\alpha_i$. For symmetric matrices $A$, $\lambda(A)$ denotes the vector of eigenvalues of $A$. $\|A\|_F = \sqrt{tr\,AA^T}$ is the Forbenius-norm of $A$. We will also use the following sets of matrices:

$$o = \{X : X^T X = XX^T = I\} \quad (7)$$
$$\varepsilon = \{X : X_e = X^T e = e\} \quad (8)$$
$$N = \{X : X \geq 0\} \quad (9)$$

Where $e$ denotes the vector with all components equal to one and $I$ the identity matrix. We note that

$$\Pi = o \cap \varepsilon \cap N \quad (10)$$

## 2. A One-Parametric Model

This problem is a generalization of unconstrained zero-one quadratic problems, zero-one quadratic knapsack problems, quadratic assignment problems and so on. It is clearly NP-hard.

Let $\nabla f(X) = 2AXB + C$ be the gradient of $f$. Concerning the information of $\nabla f(X_C)$, we can find out a better feasible solution from a linear bi-objective programming

$$\min \begin{pmatrix} tr\,\nabla f(X_C)X^T \\ tr\,-X_C X^T \end{pmatrix} \quad (11)$$
$$s.t. \quad X \in \Pi \quad (12)$$

which can be approximately solved by a parameterized LAP

$$\min tr(\alpha \nabla f(X_C) - \beta X_C)X^T \quad (13)$$
$$s.t. \quad X \in \Pi \quad (14)$$

And $\alpha \geq 0$, $\beta \geq 0$. Denote the optimal solution of (13)-(14) by $X(\alpha,\beta)$ for each fixed $(\alpha,\beta)$. It is easy to verify that $X(0,\beta)$ is the optimal solution of (5)-(6) for any $\beta > 0$. For any $\alpha > 0$ (13)-(14) can be further reduced to a one-parametric model

$$\min tr(\nabla f(X_C) - \theta X_C)X^T \quad (15)$$
$$s.t. \quad X \in \Pi \quad (16)$$

where $\theta \geq 0$. The optimal solution of (15)-(16) is denoted by $X(\theta)$.

Now we have to solve the following one-variable problem

$$\min f(X(\theta)) \quad (17)$$

$$s.t. \quad \theta \geq 0 \quad (18)$$

It is easy to verify.

**Proposition 2.1.** $f(X(\theta))$ is a piecewise constant function with respect to θ.

Illustrations can be found in Fig. 1 and Fig. 2. From Proposition 2.1, $f(X(\theta))$ is discontinuous except the trivial cases. In this article we inexactly solve it by the canonical 0.618 search method and obtain the final solution $(\theta_1, X_1)$. Generally we cannot guarantee $\theta_1$ to be optimal.

## 3. A Near-Optimum Parameter

Instead of solving the above one-parametric optimization problem, we obtain a good predestinate of the best $\theta$ in this section. Let the $(n-1)\times(n-1)$ matrix $V$ be such that

$$V^T e = 0, \quad V^T V = I_{n-1},$$

and

$$P = \left[\frac{e}{\|e\|} : V\right] \in o$$

**Theorem 3.1.** [9] Let X be $(n\times n$ and Y be $(n-1)\times(n-1)$. Suppose that X and Y satisfy

$$X = P\begin{bmatrix} 1 & 0 \\ 0 & Y \end{bmatrix} P^T \quad (19)$$

Then

$$X \in \varepsilon, \quad X \in N \Leftrightarrow VYV^T \geq -\frac{1}{n}ee^T$$
$$X \in o_n \Leftrightarrow Y \in o_{n-1}$$

Conversely, if $X \in \varepsilon$, then there is a Y such that (19) holds.

Based on the projection (19), the reduced hessen matrix of (2) is $H_r = V^T BV \otimes V^T AV$. We then have:

**Theorem 3.2.**

$$\gamma^* = \frac{(n.\,trA - e^T Ae)(n.\,trB - e^T Be)}{(n-1)^2 n^2} \quad (20)$$

Solves

$$\min \|H_r - \gamma.I\|_F \quad (21)$$

independent of the choice of V.

*Proof.* Denote the spectral decomposition of the $(n-1)\times(n-1)$ matrices $V^T AV$ and $V^T BV$ by $V^T AV = WA^A W^T$ and $V^T BV = UA^B U^T$ respectively, where W and U are orthogonal matrices, $A^A = Diag(\lambda(V^T BV))$, $A^B = Diag(\lambda(V^T BV))$.

We have

$$\|H_r - \gamma..I\|_F = \|V^T BV \otimes V^T AV - \gamma..I\|_F \quad (22)$$

$$= \|UA^B U^T \otimes WA^A W^T - \gamma..I\|_F \quad (23)$$

$$\|A^B \otimes A^A - \gamma..I\|_F \quad (24)$$

Therefore the solution of (21) is

$$\gamma^* = \frac{\sum_{i,j} A_{ii}^B A_{jj}^A}{(n-1)^2} = \frac{(\sum_i A_{ii}^B)(\sum_j A_{jj}^A)}{(n-1)^2} \quad (25)$$

Since

$$\sum_{ij} A_{ii}^A = tr V^T AV = tr VV^T A = tr(I - \frac{1}{n}ee^T) = tr A - \frac{1}{n}e^T Ae \quad (26)$$

and analogously $\sum_i A_{ii}^B = tr B - \frac{1}{n}e^T Be$, (25) is equal to (20). Obviously it is independent of the choice of V.

Practical data of A and B always have special structures as described in the following corollary.

**Corollary 3.3.** Assume $A_{ij} \geq 0, B_{ij} \geq 0$ for all $i \neq j$ and $A_{ii} = B_{ii} = 0$ =0 for all i. Then (20) reduces to

$$\gamma^* = \frac{(e^T Ae)(e^T Be)}{(n-1)^2 n^2} \geq 0. \quad (27)$$

We notice that the reduced Hessen matrix of $tr \gamma XX^T$ is $\gamma.I$ Therefore $tr \gamma^* XX^T$ is a best apporximator of $tr AXBX^T$ in the sense shown in Theorem 3.2. In other words, $f(X) = tr(AXB+C)X^T$
$= f(X_C) + tr \nabla f(X_C)(X - X_C)^T tr A(X - X_C)B(X - X_C)^T$.
can be approximated by $f(X_C) + tr\nabla f(X_C)(X - X_C)^T$
$+ tr\gamma^*(X - X_C)(X - X_C)^T tr(\nabla f(X_C) - 2\gamma^* X_C)X^T + const$ where $const = f(X_C) - tr\nabla f(X_C)X_C^T + tr\gamma^* X_C X_C^T + n\gamma^*$ is a constant. Comparing this linear function with (15), we set $\theta^* = 2\gamma^*$ and solve the corresponding LAP. This solution is denoted by $(\theta_2, X_2)$.

## 4. Numerical Results

In this section, we report our numerical experiments results on some problems from QAPLIB [10]. The infeasible solution $X_C$ was generated as follows. Firstly $r$ permutation matrices $X^1, X^2,,..., X^r$ are generated at random. We combine them to obtain $X_C$, i.e., $X_C = \sum_i^r \lambda_i X^i$, where $\lambda_i \geq 0$ and $\sum_i^r \lambda_i = 1$.

Here, we set $\lambda_i = \frac{1}{r}$ and r=2 and $[\frac{n}{2}]$ corresponding to the sparse and dense cases respectively.

In our implementation of 0.618 search method, the initial sector is [0, M] and the procedure stops if the length of the sector is less than 1. Here we set $M = \max(\theta_2, 100)$. The numerical results reported in Table 1 and Table 2 were the average values of 10 independent runs. The last three columns gave the values of $\frac{f(X_0)}{Max}, \frac{f(X_1)}{Max}$ and $\frac{f(X_2)}{Max}$ respectively, where $Max = \max(f(X_0), f(X_1), f(X_2))$. Since $f(X_0)$ is generally weak in quality and $f(X_1)$ is time-consuming, $f(X_2)$ is a good choice in practice. Furthermore, as the problem size grows, it seems that the ratio between $f(X_1)$ and $f(X_2)$ is closer due to Max is larger and M is not large enough.

It has been shown in Fig. 1 and Fig. 2 that $\theta_1$ is not guaranteed to be an optimal solution $f(X_0)$ while sometimes $\theta_2$ can be optimal while $\theta_1$ cannot.

## 5. Conclusions

We extended the most-used rounding approach to a one-parametric optimization model for the quadratic assignment problems. This model can be applied to other nonlinear assignment problems directly. We inexactly solved this model by 0.618 search method. Better choices depend on the balance between the stopping criteria and solution quality. We preestimated a near-optimum parameter rather than solving the one-parametric model for quadratic assignment problems. Our numerical experiments confirmed the efficiency. Furthermore, the prior parameter can be regarded as the initialization of the one-parametric model and several iterations could be implemented to improve it.

**Table 1** *Comparison among $f(X_0)$, $f(X_1)$ and $f(X_2)$.* $(r = 2)$.

| Problem | $f(X_0)/Max$ | $f(X_1)/Max$ | $f(X_2)/Max$ |
|---|---|---|---|
| nug 20 | 1 | 0.90 | 0.96 |
| nug 30 | 1 | 0.92 | 0.97 |
| kra 30a | 1 | 0.89 | 0.89 |
| kra 30b | 1 | 0.90 | 0.90 |
| sko 49 | 1 | 0.96 | 0.98 |
| sko90 | 1 | 0.97 | 0.98 |
| wil 100 | 1 | 0.98 | 0.99 |
| tho 150 | 1 | 0.98 | 0.98 |

**Table 2** *Comparison among $f(X_0)$, $f(X_1)$ and $f(X_2)$.* $(r = [\frac{n}{2}])$.

| Problem | $f(X_0)/Max$ | $f(X_1)/Max$ | $f(X_2)/Max$ |
|---|---|---|---|
| nug 20 | 1 | 0.88 | 0.92 |
| nug 30 | 1 | 0.90 | 0.94 |
| kra 30a | 1 | 0.89 | 0.90 |
| kra 30b | 1 | 0.89 | 0.90 |
| sko 49 | 1 | 0.96 | 0.98 |
| sko 40 | 1 | 0.96 | 0.97 |
| wil 100 | 1 | 0.98 | 0.98 |
| tho 150 | 1 | 0.97 | 0.97 |

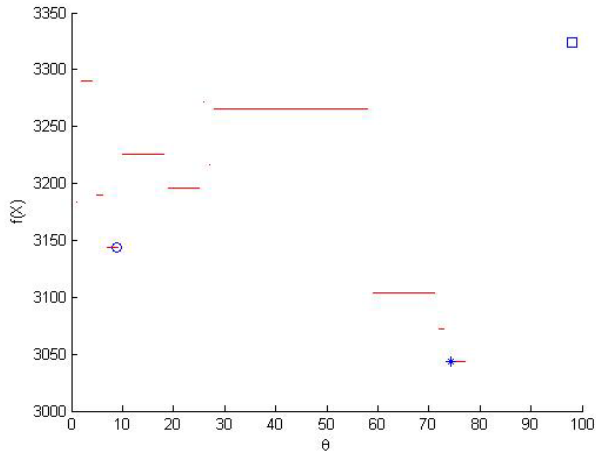

**Fig. 1**   $f(X(\theta))$  *on nug20 root problem.*

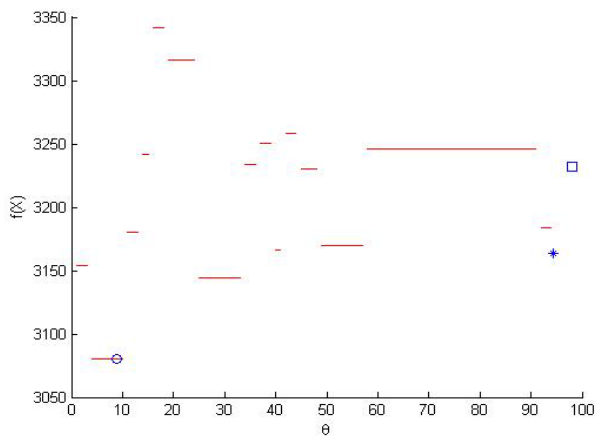

**Fig. 2**   $f(X_0), *:(\theta_1, f(X_1)), o:(\theta_2, f(X_2))$.